\documentclass[reprint,amsmath,amssymb,aps,]{revtex4-2}
\usepackage{graphicx}
\usepackage{dcolumn}
\usepackage{bm}
\usepackage{lipsum}
\usepackage[colorlinks=true,breaklinks=true,linkcolor=blue,
            citecolor=blue,urlcolor=blue]{hyperref}
\usepackage{tgtermes}
\usepackage{txfonts}
\usepackage{xcolor}
\usepackage{asymptote}
\usepackage{tikz}

\begin{document}

\preprint{APS/123-QED}

\title{Signature of inertia on light dragging in rotating plasmas}

\author{Julien Langlois}
\email{julien.langlois@laplace.univ-tlse.fr}
\author{Renaud Gueroult}
\affiliation{LAPLACE, Université de Toulouse, CNRS, INPT, UPS, 31062 Toulouse, France}

\date{\today}

\begin{abstract}
    The signature of light dragging in a rotating unmagnetized plasma is studied analytically. In contrast with previous work which focused exclusively on the drag effects arising from rigid rotation, we examine here the supplemental contribution of inertia to the rest-frame dielectric properties of a rotating medium. We reveal, for the first time, that these so far neglected contributions actually play a dominant role on light dragging in rotating unmagnetized plasmas. Besides birefringence and enhanced polarization drag, inertia is notably demonstrated to be the cause of a non-zero drag, pointing to fundamental differences between linear and angular momentum coupling. We finally discuss how, thanks to the more favourable scaling elicited here, it may be possible to observe these effects in recently proposed laser driven rotating plasmas, identifying new promising directions for experimental investigations.   
\end{abstract}

\maketitle

\textit{Introduction}.---Wave propagation in a medium is affected by the medium’s motion, leading to many fascinating physical phenomena including the Doppler effect~\cite{Doppler1842}, negative refraction~\cite{Grzegorczyk2006}, the Fresnel drag by which a moving medium drags light~\cite{fresnel1818} or optical analogs to black holes' event horizon~\cite{Philbin2008}. Propagation in moving media is also captivating in that it is intrinsically nonreciprocal~\cite{Caloz2018,Asadchy2020} as a result of a broken time-reversal symmetry~\cite{Kong2008}. Understanding the effect of motion on propagation thus holds promise both to study numerous fundamental phenomena~\cite{Potton2004} and to develop new applications~\cite{Bi2018,Nagulu2020}.

For a uniform linear motion along the beam direction, light dragging materializes as a longitudinal shift, as originally postulated by Fresnel~\cite{fresnel1818} and demonstrated by Fizeau~\cite{fizeau1860xxxii}. When the motion has a component perpendicular to the beam, light dragging shows as a deflection of the beam in this direction~\cite{jones1972fresnel,jones1975}, which leads to a lateral shift of the beam. This deflection is generally understood as the beam refraction seen by an observer in the moving frame~\cite{carusotto2003transverse}, yielding the classical ${1-1/(\bar{n}_g \bar{n})}$ drag coefficient~\cite{player1975,rogers1975}, with $\bar{n}$ and $\bar{n}_g$ the phase and group index of the wave, respectively. A direct consequence of this scaling is that drag effects are null in an unmagnetized plasma~\cite{ko1978passage}. Fundamentally this result stems from the fact that the dispersion relation for a cold unmagnetized plasma is Lorentz-invariant. This absence of drag has also been suggested to support Minkowski's formulation of momentum partitioning in a medium~\cite{arnaud1976dispersion,jones1979radiation}. 

Beyond a uniform motion, the question of the effect of an accelerated motion was naturally raised~\cite{Heer1964,Anderson1969}. In fact these questions recently regained attention with the rapid development of space-time metamaterials mimicking moving media~\cite{Huidobro2019,Caloz2020}, and more particularly with its generalization to accelerated motion~\cite{Bahrami2023}. Dragging effects due to an accelerated motion, including rotation, are generally examined by considering a Lorentz transform between the laboratory frame $\Sigma$ and the inertial frame in which the medium is instantaneously at rest $\Sigma'$~\cite{Heer1964,Anderson1969,player1976,gueroult2019}. A challenge with this method though is that it demands as inputs the constitutive relations in the instantaneous rest-frame~\cite{Anderson1969}, which, contrary to a uniform motion, can in principle no longer be considered unaffected for an accelerated motion. For a medium that is linear and isotropic at rest, and using prime and bar notations to refer to variables expressed in $\Sigma'$ and for a medium at rest, respectively, this implies determining $\epsilon'$ and $\mu'$ as function of $\bar{\epsilon}$, $\bar{\mu}$ and the velocity. For rotation this has notably been shown to lead to a gyrotropy induced by the Coriolis acceleration~\cite{shiozawa1973,shiozawa1974,langlois2023contribution}, i.e. to a tensorial permittivity $\epsilon'$. Yet, because these corrections are predicted to be small in dielectric media away from resonances~\cite{shiozawa1973,shiozawa1974}, inertial corrections to rest-frame properties have so far mostly been neglected~\cite{Mazor2019,Steinberg2023,Geva2023,Bahrami2023} (i.e. ${\epsilon'=\bar{\epsilon}}$ and ${\mu'=\bar{\mu}}$), and the classically established signature of drag effects then comes exclusively from the rigid rotation of the constitutive elements of the medium~\cite{player1976,nienhuis1992,gueroult2019}. 

In this Letter we show that inertial corrections to rest-frame properties are on the contrary essential in rotating unmagnetized plasmas, where they are found to critically modify light and polarization drag signatures. Inertia is notably responsible for non-zero drag and birefringence, uncovering fundamental differences between linear and angular momentum coupling between wave and medium. By deriving the signature of these inertial contributions, we reveal, for the first time, that they in fact dominate over the drag effects classically associated with rigid rotation. We finally underline that these enhanced drag effects may be large enough to be measured in laser driven rotating plasma experiments, paving the way for experimental demonstration. 

\textit{Azimuthal rotary drag and birefringence}.---We consider here a dielectric of length $L$ that rotates at constant angular frequency $\boldsymbol\Omega=\Omega\mathbf{\hat{e}}_z$, and we will often take this dielectric to be an unmagnetized cold collisionless fully-ionized plasma.

\begin{figure}
    \includegraphics{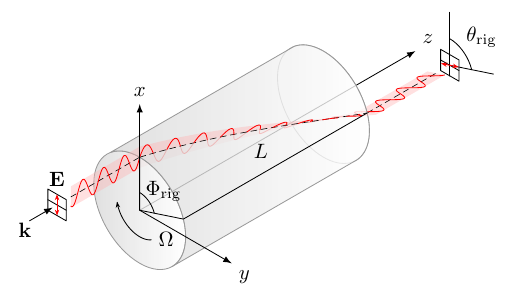}
    \caption{\label{fig:fig_1} Azimuthal rotary drag $\Phi$ (i.e. azimuthal beam displacement) and polarization drag $\theta$ (i.e. polarization rotation) of light in a rotating dielectric. For rigid rotation with ${\mathbf{k}\parallel\boldsymbol\Omega}$ in a typical dielectric one finds ${\Phi_{\textrm{rig}}=\theta_{\textrm{rig}}}$.}
\end{figure}

As first observed by Jones~\cite{jones1972fresnel,jones1975}, light propagating parallel to the rotation axis of a rotating dielectric (wave vector ${\mathbf k\parallel \boldsymbol\Omega}$) is subject to an azimuthal drag. This phenomenon, illustrated in Fig.~\ref{fig:fig_1}, has been interpreted by applying results known for a uniform transverse motion~\cite{player1975} to the case of a transverse velocity equal to the tangential velocity of the rotating medium~\cite{padgett2006polarization,leach2008aether}, yielding to first order in $\Omega/c$ a dragging angle
\begin{equation} \label{eq:drag_angle_formula_rig}
 \Phi_{\textrm{rig}} = \frac{L\Omega}{c} \left[\bar{n}_g-\frac 1{\bar{n}} \right].
\end{equation}
Besides supporting Jones' results, this analysis was central to explain image rotation~\cite{Franke-Arnold2011}, i.e. the rotation of the transverse structure of a wave propagating through a rotating dielectric, as a frequency shift between waves with opposite orbital angular momentum~\cite{gotte2007,Wisniewski-Barker2014}. Yet, by relying on the theory derived for a uniform motion~\cite{player1975}, this model only captures the effect of a rigid rotation and fails to account for the possible effect of rotation on the properties of the dielectric in $\Sigma'$, and notably of inertia induced anisotropy~\cite{shiozawa1973,shiozawa1974,langlois2023contribution}.

To quantify the effect of inertia on drag, we follow here Leach \textit{et al.}~\cite{leach2008aether}, i.e. we consider the drag induced by a transverse velocity equal to the tangential velocity of the rotating medium, but this time we take into account the modifications to the rest-frame dielectric properties induced by Coriolis and centrifugal forces~\cite{langlois2023contribution}. A critical difference, which to our knowledge had gone unnoticed thus far, is that because of the inertia induced rest-frame gyrotropy~\cite{shiozawa1973,shiozawa1974,langlois2023contribution} an incident ray will now be split in two different refracted rays. In other words the rotational motion of an isotropic medium is the source of birefringence, in contrast with the uniform linear motion for which special relativity ensures no birefringence~\cite{Ferrell1965}. Each of these rays, denoted here $(+)$ and $(-)$, is subject to a distinct transverse drag, with a deviation angle~\cite{Supplementary}
\begin{equation} \label{eq:drag_angle_formula}
 \Phi_\pm = \frac{L\Omega}{c} \left[n_{g\pm}'-\frac 1{n_{\pm}'} \right].
\end{equation}
This is a rotational analog to the transverse drag of o- and e-beams recently observed by Hogan \emph{et al.}~\cite{Hogan2023}. Also, one recovers immediately from Eq.~(\ref{eq:drag_angle_formula}) the drag angle Eq.~(\ref{eq:drag_angle_formula_rig}) previously derived for ${n_g'=\bar{n}_g}$ and ${n'=\bar{n}}$~\cite{leach2008aether}, i.e. neglecting rest-frame corrections. More generally Eq.~(\ref{eq:drag_angle_formula}) shows that the effect of rest-frame corrections on drag, and notably on beam splitting, will be negligible when 
\begin{equation} \label{eq:condition}
    n_g' n'-1 \ll 2\frac{n'}{\bar{n}}(\bar{n}_g \bar{n}-1).
\end{equation}

\begin{figure}
    \includegraphics{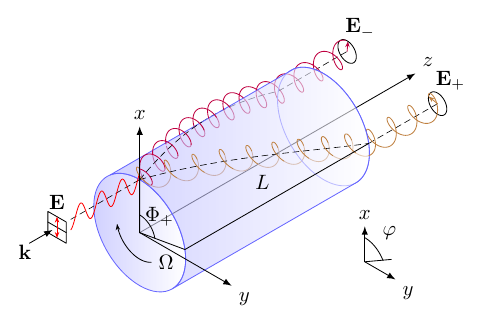}
    \caption{\label{fig:fig_2} Drag experienced by a wave incident on a rotating unmagnetized plasma. The incident beam is split into two beams due to rest-frame anisotropy. Each of these beam has a different circular polarization (right- vs left-circularly polarized) and experiences distinct drag effects $\Phi_-^{up}$ and $\Phi_+^{up}$. Polarization drag occurs when the two beams overlap.}
\end{figure}

As hinted at earlier, a striking example where Eq.~(\ref{eq:condition}) is not satisfied is an unmagnetized plasma ($up$) since in this case ${\bar{n}_g \bar{n}=1}$~\cite{ko1978passage,Supplementary}. The drag associated with rigid rotation $\Phi_{\textrm{rig}}^{up}$ is thus zero, and any contribution to azimuthal drag must stem entirely from inertia. Specifically, one finds 
\begin{equation}\label{Eq:indexes}
(n_g'n')_\pm^{up} \sim 1 \pm \frac{\Omega \omega_{pe}^2}{\omega^3}
\end{equation}
for ${\omega\geq\omega_{pe} \gg \Omega}$ with $\omega_{pe}$ the plasma frequency. From Eq.~\eqref{Eq:indexes}, we remark as illustrated in Fig.~\ref{fig:fig_2} that the two rays are dragged in opposite directions. Specifically, since ${(n_g'n')_-^{up}<1}$, the $(-)$ ray experiences anomalous drag~\cite{banerjee2022anomalous}, i.e. is being dragged in the direction opposite to the medium motion. Note that this behaviour, which again is here entirely due to inertia, differs from that in a birefringent glass rod where o- and e-beams are dragged in the same direction~\cite{Hogan2023}. 

\begin{figure*}
    \includegraphics[width=\linewidth]{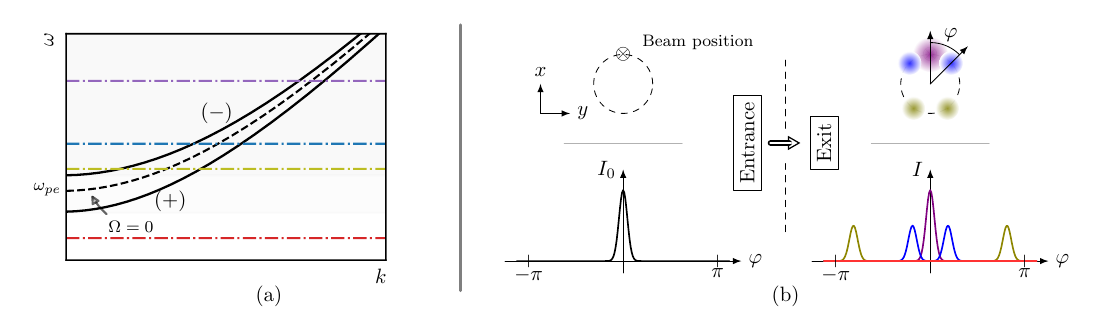}
    \caption{\label{fig:fig_3} Dispersion diagram in the laboratory frame $\Sigma$ for the two modes ($+$) and ($-$) propagating along the rotation axis of an unmagnetized plasma (a) and illustration of the frequency-dependent at azimuthal drag experienced by an incident pulse centered at ${\varphi=0}$ (b). The different colors for the transmitted signal represent components of different frequencies as shown on the dispersion diagram. Frequencies close to the cutoff $\omega_{pe}$ are strongly rotated since ${\Phi_{\pm}^{up}\rightarrow\infty}$ at the cutoff, whereas frequencies much above the cutoff are barely affected.}
\end{figure*}

Quantifying this effect (see Ref.~\cite{Supplementary}), we get from Eq.~(\ref{Eq:indexes})
\begin{equation} \label{eq:azimuthal_drag}
    \Phi_\pm^{up}=\pm\frac{L\Omega^2\omega_{pe}^2}{c\omega^3}\left[1-\frac{\omega_{pe}^2}{\omega^2}\right]^{-1/2}
\end{equation}
to leading order in $\Omega/\omega$. This shows that inertia corrections will be small for wave frequencies much above the plasma frequency. On the other hand the two transmitted rays will exhibit a strong azimuthal separation ${\Phi_\pm^{up}\rightarrow\pm\infty}$ just above the cutoff $\omega_{pe}$. This is because the associated linear drag angle reaches the critical angle $\pm\pi/2$ at the cutoff, or said differently because the transverse drag then becomes infinite (see Ref.~\cite{Supplementary}). Also, because ${\Phi_\pm^{up}\propto \pm [\omega-\omega_{pe}]^{-1/2}}$ near the cutoff, different frequency components of a pulse will be separated azimuthally as shown in Fig.~\ref{fig:fig_3}. Rotating unmagnetized plasmas thus offer opportunities to develop tuneable spectrometers in which both the operating range and the resolving power can be controlled through $\Omega$ and the plasma density.

Note also and importantly that while Eq.~\eqref{eq:azimuthal_drag} shows that ${\Phi_-^{up}=-\Phi_+^{up}}$ when ${\Omega/\omega\ll1}$, the drag angles in fact differ as $\Omega/\omega$ grows. This suggests that angular momentum is transferred from the plasma to the wave, in contrast to the result known for a uniform linear motion~\cite{ko1978passage}. The inertia driven rest-frame gyrotropy would then enable coupling between wave and plasma angular momentum, similarly to what is achieved by a magnetic field~\cite{Rax2023}. This is important in that it supports the idea that the rotation of an unmagnetized plasma can conversely be sustained through wave momentum input, whereas it would not be possible without the inertia corrections exposed here. 

\textit{Polarization drag}.---A second manifestation of rotation closely related to azimuthal drag is the polarization drag first conjectured by Thomson~\cite{thomson1885} and then derived by Fermi~\cite{fermi1923}, i.e. the rotation of the plane of polarization of a linearly polarized wave propagating along the rotation axis of a rotating dielectric (see Fig.~\ref{fig:fig_1}). This phenomenon, also first observed by Jones~\cite{jones1976}, has been interpreted as a mechanically induced circular birefringence~\cite{player1976}, identifying to first order in $\Omega/c$ the same scaling 
\begin{equation} \label{eq:polarization_drag_angle_formula_rig}
 \theta_{\textrm{rig}} = \frac{L\Omega}{c} \left[\bar{n}_g-\frac 1{\bar{n}} \right]
\end{equation}
as that uncovered for the azimuthal drag in Eq.~\eqref{eq:drag_angle_formula_rig}. Yet, similarly Eq.~\eqref{eq:drag_angle_formula_rig}, this result only accounts for rigid rotation. This shortcoming was in fact pointed out by Baranova and Zel'dovich~\cite{baranova1979}, who concluded that the contribution of inertial corrections to polarization rotation were second order compared to the solid rotation result for the glass rod and optical frequencies used by Jones~\cite{jones1976}. However this conclusion may not hold in different conditions, notably in an unmagnetized rotating plasma for which Eq.~\eqref{eq:polarization_drag_angle_formula_rig} predicts ${\theta_{\textrm{rig}}=0}$ since again ${\bar{n}_g\bar{n}=1}$. In addition,  Baranova and Zel'dovich did not consider the inertia induced beam splitting discussed above.

When inertial corrections are neglected, the medium is isotropic in $\Sigma'$. Refraction thus leads to a single ray, which is linearly polarized. That an observer in the lab-frame interprets this beam, as suggested by Player~\cite{player1976}, as the sum of two circularly polarized modes with indices ${n_{\textrm L}\neq n_{\textrm R}}$ does not change the fact that it is a single ray. In contrast, since this same medium is anisotropic in $\Sigma'$ when accounting for inertia corrections, one must then consider two refracted rays. Also, because of the change of frame of reference (${\mathbf{k}\cdot\mathbf{\hat{e}}_{x,y}=0\rightarrow\mathbf{k'}\cdot\mathbf{\hat{e}}_{x,y}\neq 0}$), propagation is oblique in $\Sigma'$, and the eigenmodes thus have complex polarizations. Yet, one verifies that these two modes correspond to lowest order in $v/c$ to circularly polarized eigenmodes in $\Sigma$~\cite{Hebenstreit1979}. From this we conclude as shown in Fig.~\ref{fig:fig_2} that the two refracted rays as seen in the lab-frame exhibit opposite circular polarization, and that polarization rotation in fact only occurs when these two rays overlap.

Unlike azimuthal drag which required modelling the tangential velocity as a linear motion, the effect of rest-frame modifications due to inertia to polarization drag can be derived directly. For that we use here that for an incident wave parallel to the rotation axis the refracted wave vectors $\mathbf{k}_{\pm}$ are also aligned with the rotation axis~\cite{Supplementary}. Using this finding, the wave equation known for rotating gyrotropic media whose axis is aligned with $\boldsymbol\Omega$~\cite{gueroult2019,gueroult2020} and the rest-frame susceptibility tensor of an unmagnetized rotating plasma~\cite{Supplementary}, we obtain an expression for the wave index difference ${n_{\textrm L}-n_{\textrm R}}$ as seen from the lab-frame $\Sigma$. For ${\omega\geq\omega_{pe} \gg \Omega}$ this yields a polarization rotation~\cite{Supplementary}
\begin{equation} \label{eq:polarization_drag_angle_short}
    \theta^{up} = \frac{L\Omega}{c}\left[\bar n-\frac {1}{\bar n} \right] + \mathcal O\left(\left[\frac \Omega\omega\right]^3 \right ).
\end{equation}
Importantly, we uncover here that the contribution of inertia corrections to polarization rotation are first order in $\Omega/c$, whereas the rigid rotation contribution is at most third order in $\Omega/c$. This demonstrates that inertia is the dominant contribution to polarization drag in rotating unmagnetized plasmas, and thus critical to model this effect.

Assuming ${\omega\gg\omega_{pe}}$, one gets from Eq.~(\ref{eq:polarization_drag_angle_short})
\begin{equation}
    \theta^{up} \sim -\frac{L\Omega}{c} \frac{\omega_{pe}^2}{\omega^2}
\end{equation}
to leading order in $\Omega/c$. This is precisely the Faraday rotation expected for a magnetic field ${B^* = -{2m_e \Omega}/{e}}$ with $m_e$ and $-e$ the electron mass and charge, supporting the Faraday-Coriolis equivalence suggested by Baranova and Zel'dovich~\cite{baranova1979}. Closer to the cutoff ${\omega\sim\omega_{pe}}$, a Puiseux expansion gives the even more favorable scaling
\begin{equation} \label{eq:max_polarization_drag}
    \theta^{up}(\omega_{pe}) \sim -\frac{L\Omega}{c} \sqrt{\frac{\omega_{pe}}{\Omega}}.
\end{equation}

\textit{Opportunities for experimental evidence}.---Having highlighted the critical role of inertia on light dragging in a rotating unmagnetized plasma, we now discuss how the singular signature of these effects could be revealed in experiments. 

Starting with azimuthal drag, the prefactor ${L\Omega^2/(c\omega_{pe})}$ in Eq.~(\ref{eq:azimuthal_drag}) suggests that this effect will generally be small. Indeed, one expects ${L\Omega/c\ll1}$ in most experiments, and ${\Omega\ll\omega_{pe}}$ is assumed in the model derived here. Achieving ${\Phi_\pm^{up}=\mathcal{O}(1)}$ thus only appears possible in the immediate proximity of the cutoff where ${\bar{n}\rightarrow 0}$. On the other hand, smallness of $\Phi_\pm^{up}$ appears desirable to measure polarization drag. Strong azimuthal drag will indeed lead to a strong beam splitting, which, short of particular conditions where the two refracted beams are spatially recombined (e.g. for ${\Phi_+^{up}=-\Phi_-^{up}=\pi}$), will preclude the observation of polarization drag. Looking at Eq.~(\ref{eq:max_polarization_drag}), we note conversely that polarization drag $\theta^{up}(\omega_{pe})$ scales as ${(\omega_{pe}/\Omega)^{1/2}\gg1}$, which creates opportunities. 

\begin{figure}
    \includegraphics[]{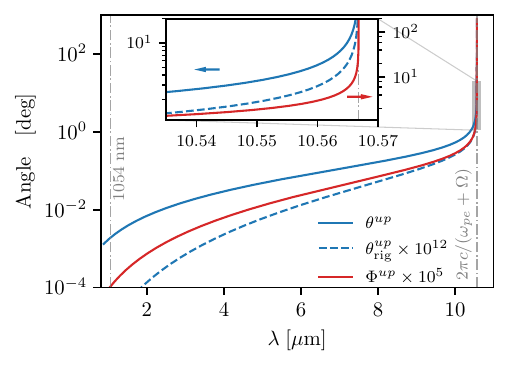}
    \caption{\label{Fig:HEDP} Azimuthal drag angle $\Phi^{up}$ ($\times10^5$) and polarization drag angle $\theta^{up}$ dependence on the wavelength $\lambda$ for the conditions in laser driven rotating unmagnetized plasmas. The uncorrected polarization drag angle $\theta_{\textrm{rig}}^{up}$ ($\times10^{12}$) is shown for comparison. The inset highlights the behaviour near the cutoff ${2\pi c/(\omega_{pe}+\Omega)}$ where drag effects are strongly enhanced. ${n_e=10^{25}\text{ m}^{-3}}$, ${\Omega=10^9\text{ rad.s}^{-1}}$ and ${L=10^{-3}\text{ m}}$.}
\end{figure}

To illustrate these opportunities we consider here as an example the rotating unmagnetized plasma produced by merging plasma outflows created by the laser irradiation of purposely shaped targets~\cite{ryutov2011using}. This rotation drive scheme was recently successfully demonstrated on the Omega laser facility, creating high density rotating plasmas with ${n_e=10^{25} \text{ m}^{-3}}$ and ${\Omega=10^9 \text{ rad.s}^{-1}}$ about a mm long~\cite{suzuki2021laser,SuzukiVidal2022}. While as shown in Fig.~\ref{Fig:HEDP} the polarization rotation predicted in these conditions is small for the laser wavelength ${\lambda = 351-1053~\text{ nm}}$ typically used, the $\omega^{-2}$ scaling leads to a polarization rotation angle of nearly ${\theta^{up}\sim-80\text{ deg}}$ near the mid-infrared cutoff ${\omega_{pe}\sim 1.8~10^{14}~\text{ rad.s}^{-1}}$ (${\lambda_{pe}\sim 10.5~\mu\text{m}}$). This is orders of magnitude larger (more than $10^{11}$ times for the parameters used here) than the value $\theta_{\textrm{rig}}^{up}$ obtained without inertia corrections. This large separation, which is consistent with the $\Omega/\omega$ and $(\Omega/\omega)^3$ scalings predicted in Eq.~\eqref{eq:polarization_drag_angle_short}, should allow for an unequivocal demonstration of inertia corrections. One also verifies in Fig.~\ref{Fig:HEDP} that the drag angle remains in these same conditions much smaller than the polarization drag angle. This ensures that the two refracted beams will mostly overlap, leading to polarization rotation. While the mid-infrared diagnostics required for this measurement are not presently available on Omega - and in general not often found in laboratory plasma experiments, we note here that mid-infrared polarimetry has in contrast received considerable attention in recent years in astrophysics~\cite{Aitken2004,Telesco2022}. The tools developed in this context, combined with the availability of tunable mid-infrared quantum cascade lasers~\cite{Yao2012,Botez2023}, could be applied in laboratory plasma experiments. {It is important to note however that the predictions given here rely on a homogeneous cold plasma model, which is oversimplified. The possibility to observe inertia's signature on polarization will thus have to be confirmed for more realistic conditions, accounting notably for the possible effects of temperature, inhomogeneities and absorption.}

Lastly, note that because this effect is non-reciprocal, one way to lighten the constraints to measure this effect is to use multi-path to get an $n$-fold increase in the drag angle, as in fact done originally by Jones~\cite{jones1976}. Although this approach may not lend itself well to the high energy density platform discussed here due to the short timescale at play, it could in principle be used on a steady-state mechanically driven rotating plasma column.

\textit{Conclusions}.---We showed that the effect associated with inertia corrections to the rest-frame properties of a medium in non-uniform motion, which have so far mostly been neglected in the literature, can have a critical impact on wave dynamics.

Considering that a dielectric that is isotropic at rest exhibits gyrotropic properties when rotating, we showed that rotation is the source of birefringence which affects the classical light dragging picture. We then exposed, for the first time, how these effects are particularly important in the case of a cold rotating unmagnetized plasma. We specifically demonstrated how in this case classical models neglecting inertia corrections predict zero drag, whereas in fact inertia corrections lead to a finite azimuthal drag. This is important in that it suggests that wave momentum could be used to sustain plasma rotation but not uniform translation, pointing to fundamental differences between linear and angular momentum coupling.

Besides finite azimuthal drag, we uncovered that inertia corrections to rest-frame properties are also responsible for an augmented polarization drag in rotating unmagnetized plasmas, with the so far neglected inertia effects now dominating over the polarization drag classically associated with rigid rotation. We finally showed that these enhanced effects may be large enough to be measured in recently proposed laser driven rotating plasma experiments, paving the way for an experimental demonstration of the effect of inertia on the dielectric properties of rotating media and on propagation.

Lastly, while the demonstration of the importance of inertia effects on light dragging reported here was made possible thanks to the the simplicity of the dielectric model of a cold unmagnetized plasma, our work more generally underlines the importance of carefully assessing how rest-frame properties are affected by an accelerated motion, and how these modifications can carry over to wave dynamics. This is particularly relevant for the numerous applications relying on high-precision predictions of wave dynamics in rotating environments, such as energy deposition in a tokamak~\cite{Slief2023} for magnetic confinement fusion or the modeling of atmospheric occultations~\cite{Bourgoin2021} in astrophysics. In exhibiting, as shown here, distinct and measurable differences compared to the zero light drag and weak polarization drag predicted when considering only rigid body rotation, unmagnetized plasmas offer a rare platform to study these effects. 

\begin{acknowledgments}
    This work was supported by the French Agence Nationale de la Recherche (ANR), under grant ANR-21-CE30-0002 (project WaRP). JL acknowledges the support of ENS Paris-Saclay through its Doctoral Grant Program.

    The authors would like to thank Dr. Vicente Valenzuela-Villaseca, Dr. Francisco Suzuki-Vidal and Dr. Joao Santos for constructive discussions and for pointing us to the laser driven high density unmagnetized rotating plasma experiments, as well as Aymeric Braud for stimulating discussions.
\end{acknowledgments}



%


\end{document}


\preprint{APS/123-QED}

\title{Signature of inertia on light dragging in rotating plasmas:\\supplemental material}

\author{Julien Langlois}
\email{julien.langlois@laplace.univ-tlse.fr}
\author{Renaud Gueroult}
\affiliation{LAPLACE, Université de Toulouse, CNRS, INPT, UPS, 31062 Toulouse, France}

\date{\today}

\maketitle

In this supplemental material we provide additional details on light dragging in a rotating unmagnetized plasma. First we recall in Sec.~\ref{sec:appendixA} how to obtain readily from previous contributions the susceptibility tensor of a rotating cold unmagnetized plasma in its rest-frame. Then we derive in Sec.~\ref{sec:appendixB} the full expression for  azimuthal drag accounting for inertia corrections. Finally we show in Sec.~\ref{sec:appendixC} how past work can be used to obtain a formula for polarization drag accounting for inertia corrections, discussing in passing the dispersion of the two modes propagating along the rotation axis.

\section{Rest-frame dielectric susceptibility \\ of the rotating unmagnetized cold plasma} \label{sec:appendixA}

The dielectric susceptibility tensor of an unmagnetized cold plasma at rest is isotropic and writes
\begin{equation} \label{eq:isotropic_tensor}
    \underline{\bar{\boldsymbol\chi}} = \bar\chi_\parallel \underline{\mathbf{I}} \quad\text{where}\quad \bar\chi_\parallel(\omega) = -\sum_s \frac{\omega_{ps}^2}{\omega^2}
\end{equation}
with $\omega_{ps}$ the plasma frequency of species $s$. From Eq.~\eqref{eq:isotropic_tensor} one gets
\begin{equation}
\bar{n} = \sqrt{1+\bar\chi_\parallel }=\sqrt{1-\sum_s \frac{\omega_{ps}^2}{\omega^2}}
\end{equation}
and
\begin{align}
\bar{n}_g & = \bar{n}+\omega\frac{d\bar{n}}{d\omega}\nonumber\\
& = \frac{1}{\bar{n}}
\end{align}
so that in an an unmagnetized cold plasma at rest
\begin{equation}
\bar{n}_g\bar{n}=1.
\end{equation}

Now, considering the corrections to the rest-frame dielectric properties for a non-relativistic rotating medium derived by Langlois and Gueroult~\cite{langlois2023contribution}, the rest-frame dielectric susceptibility tensor for a rotating cold unmagnetized plasma can be written as the Hermitian tensor 
\begin{equation} \label{eq:hermitian_tensor}
    \underline{\boldsymbol\chi}' = 
    \begin{pmatrix}
        \chi_\perp' & -i\chi_\times' & 0\\ 
        i\chi_\times' & \chi_\perp' & 0\\ 
        0 & 0 & \chi_\parallel'
    \end{pmatrix}
\end{equation}
where
\begin{subequations} \label{eq:tensor_components}
\begin{align}
    \chi_\parallel'(\omega') &= -\sum_s \frac{\omega_{ps}^2}{\omega'^2} \sim -\frac{\omega_{pe}^2}{\omega'^2}, \\ 
    \chi_\perp'(\omega') &= \frac{1+(\Omega/\omega')^2}{\left[1-(\Omega/\omega')^2\right]^2}~\chi_\parallel'(\omega'), \\
    \chi_\times'(\omega') &= \frac{-2\Omega/\omega'}{\left[1-(\Omega/\omega')^2\right]^2}~\chi_\parallel'(\omega').
\end{align}
\end{subequations}
One notices in particular that rotation introduces gyrotropy, as previously noted by Shiozawa~\cite{shiozawa1973,shiozawa1974}.

\section{Derivation of the azimuthal rotary drag} \label{sec:appendixB}

\begin{figure}
    \includegraphics{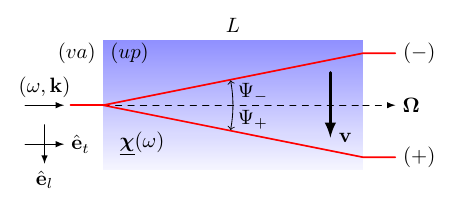}
    \caption{\label{fig:fig_1_supp} Refraction and transverse drag experienced by a beam in normal incidence on a medium in uniform linear motion as seen from the laboratory frame $\Sigma$, for a medium with the Hermitian dielectric tensor Eq.~\eqref{eq:hermitian_tensor} in its rest-frame. The beam is observed to be split in two modes ($+$) and ($-$), each exhibiting a different drag.}
\end{figure}

Consider a beam incident on an unmagnetized plasma ($up$) in uniform linear motion with velocity $\mathbf{v}$, coming in the lab-frame from vacuum ($va$) in the direction perpendicular to the plasma motion (see Fig.~\ref{fig:fig_1_supp}). The incident wave's quadrivector ${[\omega,\mathbf n^{va}]}$
where $\omega$ and $\mathbf n$ are respectively the frequency and the optical index in $\Sigma$, is rewritten in the rest frame $\Sigma'$ according to the Lorentz transforms neglecting relativistic effects
\begin{equation} \label{eq:lorentz_transforms}
    \omega' = \omega, \qquad (n_l^{va})' = -\frac{v}{c}, \qquad (n_t^{va})'=n_t^{va},
\end{equation}
where $l$ and $t$ index refer to longitudinal and transverse components with respect to motion. Now in $\Sigma'$ where the plasma is at rest we can apply Snell's law to obtain the longitudinal part of the optical index in the plasma:
\begin{equation} \label{eq:snell}
    (n_l^{up})' = (n_l^{va})'.
\end{equation}
The expression of the transverse component $(n_t^{up})'$ on the other hand depends on the form of the considered dielectric tensor, for a given wave equation. Assuming that the dielectric perturbations follow the classical ${|(n'^2-1)\underline{\mathbf I}-\mathbf n' \otimes \mathbf n'-\underline{\boldsymbol\chi}'|=0}$ dispersion relation for a medium with the Hermitian dielectric susceptibility given in Eq.~\eqref{eq:hermitian_tensor}, the problem amounts to solving an Appleton-Hartree equation for $(n_t^{up})'$. Following for instance Bittencourt~\cite{bittencourt2004fundamentals}, one finds using Eqs.~\eqref{eq:lorentz_transforms}-\eqref{eq:snell} two solutions, denoted here with index ($+$) and ($-$),
\begin{equation}\label{eq:appleton}
    (n_t^{up})_{\pm}'=\left[\frac{1}{2A}\left(B \pm \sqrt{B^2-4AC}\right)\right]^{1/2}
\end{equation}
with 
\begin{subequations} \label{eq:appleton_coeff}
\begin{align}
    A &= 1 + \chi_\parallel', \\
    B &= 2(1 + \chi_\parallel')(1 + \chi_\perp')-\frac{v^2}{c^2}(2+\chi_\parallel'+\chi_\perp'), \\
    C &= \left[(1+\chi_\perp')\frac{v^2}{c^2}-(1+\chi_\perp')^2+\chi_\times'^2\right]\left[\frac{v^2}{c^2}-(1+\chi_\parallel')\right],
\end{align}
\end{subequations}
and where all susceptibility components in Eqs.~\eqref{eq:appleton_coeff} are implicitly functions of the rest-frame frequency $\omega'$. The transverse drag $\Psi_\pm$ associated with each refractive index ${n_{\pm}'=[n_l'^2+n_{t_{\pm}}'^2]^{1/2}}$ is then obtained from Player's non-relativistic formula~\cite{player1975}
\begin{equation}
    \tan\Psi_{\pm} = \frac{v}{c} \left[n_{g_{\pm}}'-\frac 1{n_{\pm}'} \right],
\end{equation}
with the rest-frame group index ${n_{g_{\pm}}'=n_{\pm}'+\omega' dn_{\pm}'/d\omega'}$. 

Now, considering a transverse velocity equal to the tangential velocity of the rotating medium ${v=R\Omega}$, the azimuthal rotary drag $\Phi$ as defined in the Letter is found from the trigonometric relation ${\Phi=(L/R) \tan\Psi}$, giving
\begin{equation} \label{eq:drag_angle_formula}
    \Phi = \frac{L\Omega}{c} \left[n_{g}'-\frac 1{n'} \right].
\end{equation}
Injecting Eqs.~\eqref{eq:lorentz_transforms}-\eqref{eq:appleton} in Eq.~\eqref{eq:drag_angle_formula} with the tensor components of Eq.~\eqref{eq:tensor_components}, one gets for the two modes
\begin{align} \label{eq:azimuthal_rotary_drag_angle}
    \Phi_\pm^{up} = & \underbrace{\pm\frac{L \Omega}{c}\left\{ \frac{\Omega}{\omega}\left[\bar n-\frac {1}{\bar n} \right]+\mathcal O\left(\left[\frac \Omega\omega\right]^3 \right )\right \}}_{\Phi_\textrm{iner}^{up}} + \underbrace{\frac{L\Omega}{c}\left[\bar n_g-\frac {1}{\bar n} \right]}_{\Phi_\textrm{rig}^{up}=0}\nonumber\\
    = & \pm\frac{L\Omega^2\omega_{pe}^2}{c\omega^3}\left[1-\frac{\omega_{pe}^2}{\omega^2}\right]^{-1/2}+\mathcal O\left(\left[\frac \Omega\omega\right]^3\right).
\end{align}

Note for high frequency waves such that ${\omega \gg \omega_{pe}}$ this last result is consistent with the transverse drag derived for a magnetized plasma in which the magnetic field is perpendicular to the motion~\cite{meyer1980high}. This match is the direct consequence that, as predicted by Larmor's theorem~\cite{larmor1900,brillouin1945}, Lorentz and Coriolis forces play the same role on electron dynamics. High-frequency drag in a rotating unmagnetized plasma thus has the same signature as a magnetized plasma in uniform linear motion.

We note also here that by applying the inverse Lorentz transform to evaluate the optical index of ($+$) and ($-$) modes in $\Sigma$ one finds ${n_t^{up}=(n_t^{up})'}$ and ${n_l^{up}=0}$. As a result for an incident wave in normal incidence the wave vector of the two refracted waves remains perpendicular to the medium's motion, that is aligned with the wave vector of the incident wave.

\section{Derivation of the polarization drag} \label{sec:appendixC}

\begin{figure}
    \includegraphics[width=8cm]{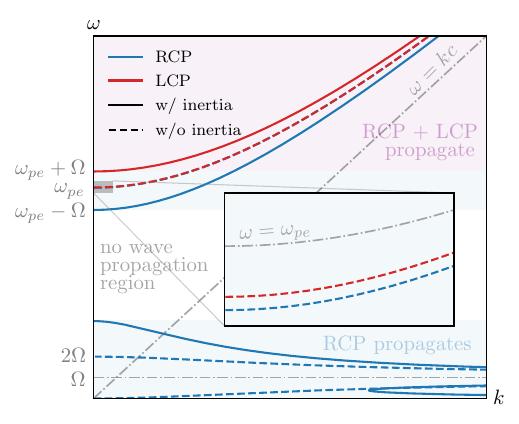}
    \caption{\label{fig:fig_2_supp} Dispersion diagram in the lab frame $\Sigma$ for the R-circularly polarized (RCP) and L-circularly polarized (LCP) modes derived in Eq.~\eqref{eq:CP_modes}. Solid lines indicate the mode dispersion accounting for inertia corrections, whereas the dashed lines represent the mode dispersion without inertia corrections to the rest-frame dielectric properties. Expressions for the cutoffs are given in Eqs.~\eqref{Eq:cutoff_full} and \eqref{Eq:cutoff_ni}.}
\end{figure}

Polarization rotation occurs when a linearly polarized wave can be decomposed into right ($\textrm R$) and left ($\textrm L$) circularly polarized eigenmodes with different optical indices, $n_\textrm{R}$ and $n_\textrm{L}$ respectively~\cite{fowles1989,barron2009}. This index difference introduces a phase shift between the two eigenmodes such that the polarisation is rotated, after a distance $L$ in the medium, by the angle
\begin{equation} \label{eq:polarization_rotation_angle}
    \theta = \frac{L\omega}{2c} \Big[n_\textrm{L} - n_\textrm{R} \Big].
\end{equation}
Meanwhile, Gueroult \textit{et al.}~\cite{gueroult2019} showed that the two modes for propagation along the rotation axis of a rotating gyrotropic medium are circularly polarized with indexes
\begin{equation} \label{eq:general_dispersion_relation}
	n_\textrm{R/L}^2 = 1 + \chi_\perp’(\omega’) \pm \chi_\times’(\omega’)
	- \frac{\Omega}\omega\Big[\chi_\times’(\omega’)\pm \chi_\perp’(\omega’)\pm \chi_{\parallel}'(\omega’)\Big]
\end{equation}
{with $\omega'=\omega\mp\Omega$ the Doppler shifted frequency. }Plugging the susceptibility tensor components of the unmagnetized cold plasma corrected for inertia effects given in Eq.~\eqref{eq:tensor_components}, it comes
\begin{equation} \label{eq:CP_modes}
    n_\textrm{R/L}^{up} = \left[1 - \left( \frac{\omega'^3}{\omega^3} \mp \frac{\Omega}{\omega}\right) \frac{\omega_{pe}^2}{\omega'^2}\right ]^{1/2}.
\end{equation}
The dispersion diagram for these two modes is shown in Fig.~\ref{fig:fig_2_supp}. From \eqref{eq:CP_modes} one finds for each of these modes a cutoff frequency near the plasma frequency $\omega_{pe}$ (i.e. the cutoff frequency for a plasma at rest) given by
\begin{equation}
    \omega_\textrm{R/L}^{up} \sim \omega_{pe} \mp \Omega.
    \label{Eq:cutoff_full}
\end{equation}
Note that neglecting inertial effects in the plasma's dielectric response---that is considering Eq.~\eqref{eq:isotropic_tensor} rather than Eq.~\eqref{eq:tensor_components} in Eq.~\eqref{eq:general_dispersion_relation}---gives instead the cutoffs
\begin{equation}
    \omega_\textrm{R/L}^{up} \sim \omega_{pe} - \frac{\Omega^2}{2\omega_{pe}}\left(1\pm\frac{2\Omega}{\omega_{pe}} \right ).
    \label{Eq:cutoff_ni}
\end{equation}

Finally, injecting Eq.~\eqref{eq:CP_modes} into Eq.~\eqref{eq:polarization_rotation_angle}, one gets the polarization rotation angle above the cutoffs
\begin{align} \label{eq:polarization_drag_angle}
    \theta^{up} &= \underbrace{\frac{L\Omega}{c}\left[\bar n-\frac {1}{\bar n} \right]}_{\theta_\textrm{iner}^{up}} + \underbrace{\left\{\frac{L\Omega }{c}\left[\bar n_g-\frac {1}{\bar n} \right]+ \mathcal O\left(\left[\frac \Omega\omega\right]^3 \right )\right\}}_{\theta_\textrm{rig}^{up}=0+\mathcal O([\Omega/\omega]^3)} \\
    = & -\frac{L\Omega\omega_{pe}^2}{c\omega^2}\left[1-\frac{\omega_{pe}^2}{\omega^2}\right]^{-1/2}+\mathcal O\left(\left[\frac \Omega\omega\right]^3\right).
\end{align}
The rigid-body rotation contribution $\theta_\textrm{rig}^{up}$ can be determined simply by neglecting inertial effects in the plasma's dielectric response, considering again Eq.~\eqref{eq:isotropic_tensor} rather than Eq.~\eqref{eq:tensor_components} in Eq.~\eqref{eq:general_dispersion_relation}.

%